\documentclass[%
preprint,
aps,
prl
]{revtex4-2}

\usepackage{graphicx}% Include figure files
\usepackage{dcolumn}% Align table columns on decimal point
\usepackage{bm}% bold math
\begin{document}

\preprint{APS/123-QED}

\title{Response of negative ion beamlet to RF field in beam extraction region}% Force line breaks with \\

\author{Kenichi Nagaoka}
%\author{Taiga Hamajima}%
 \email{nagaoka@nifs.ac.jp}
 \author{Haruhisa Nakano}
\affiliation{%
 Natioinal Institute for Fusion Science, Oroshi, Toki, 509-5292 Japan\\
 Nagoya University, Furocho, Chikusa, Nagoya 464-8602 Japan  %\textbackslash\textbackslash
}%

% \collaboration{MUSO Collaboration}%\noaffiliation

\author{Taiga Hamajima}%
 \altaffiliation{Present affiliation is Denso cooperation.}
\affiliation{%
 Nagoya University, Furocho, Chikusa, Nagoya 464-8602 Japan  %\textbackslash\textbackslash
}%

\author{Ryoya Nakamoto}%
\altaffiliation{Present affiliation is IHI cooperation.}
\affiliation{%
 Nagaoka University of Technology, Nagaoka, NIigata 940-2188 Japan  %\textbackslash\textbackslash
}%
 
\author{Katsuyoshi Tsumori}
\author{Masaki Osakabe}
\affiliation{%
 Natioinal Institute for Fusion Science, Oroshi, Toki, 509-5292 Japan % \\
 The Institute for Advanced Science, SOKENDAI, Oroshi, Toki, 509-5292 Japan% \textbackslash\textbackslash
}%

\author{Katsunori Ikeda}
\author{Masashi Kisaki}
\altaffiliation{Present affiliation is National Institute for Quantum Science and Technology.}
\affiliation{%
 Natioinal Institute for Fusion Science, Oroshi, Toki, 509-5292 Japan % \\
}%

\author{Kenji Miyamoto}
\affiliation{%
 Naruto University for Education, Naruto 772-8502 Japan %\\
}%

\author{Kazunori Takahashi}
\affiliation{%
 Tohoku Univeristy, Sendai, 980-8579 Japan %\\
}%

\author{Ursel Fantz}
\affiliation{%
 Max-Planck Institute for Plasma Physics, Boltzmannstr. 2, 85748 Garching, Germany %\\
}%


\date{\today}% It is always \today, today,

\begin{abstract}
Beam-focusing characteristics of negative ion beams have been experimentally investigated 
with a superimposition of a controlled perturbation of RF field in a filament-arc discharge negative ion source. 
Oscillations of a negative-ion beamlet width and axis responding to the RF perturbation were observed, 
which may be a cause of the larger beam divergence angle of the RF negative ion source for ITER.
It is pointed out that the oscillation of the beamlet width depends on the perveance and on the RF frequency such that the oscillation is suppressed at perveance-matched conditions and at low RF frequency.
\begin{description}
\item[Keywords]
sheath with negative ion, RF negative ion source, perveance
\end{description}
\end{abstract}

\maketitle

%\tableofcontents

%%%%%%%%%%%%%%%%%%%%%%%%%%%%%%%%%%%%%%%%%%%%%%%%%%%%%%%%%
%\section{\label{sec:level1}Introduction}
%%%%%%%%%%%%%%%%%%%%%%%%%%%%%%%%%%%%%%%%%%%%%%%%%%%%%%%%%
Negative hydrogen ions have a large cross-section for charge exchange in a high-energy regime, 
and high-energy negative ion beams have been utilized in high-energy-particle physics experiments, 
magnetically confined fusion experiments, small-size accelerators for medical applications, etc.
In order to expand their applications widely and to make breakthroughs in scientific research projects, 
further improvements in negative ion source performances 
such as long pulse operation capability, higher beam current density, better beam focusing, etc
are going on.

Recently, it has been recognized that beam divergence of the RF negative ion source for ITER [1-8] is 
larger than that of filament-arc negative ion sources,  
while it can almost satisfy the requirement for long pulse and high beam current density operation. 
The beam divergence of RF negative ion source is 9-12 mrad at an acceleration voltage of 50 kV [9-10], 
and that of the filament-arc source is 5 mrad at almost the same acceleration voltage. 
While the beam divergence may improve at 1 MeV acceleration [11],
improvement of beam divergence is an important urgent issue for 
the development of the RF negative ion source for ITER [12]. 

Behaviors of negative ion sheath near the beam extraction region play an important role in negative ion beam focusing 
because the negative ion sheath boundary, the so-called ''plasma meniscus,'' is an electrostatic lens 
at the first stage of negative ion beam acceleration. 
However, a physics model of negative ion sheath formation has not been established yet. 
Therefore, fundamental beam-focusing characteristics have been investigated with experiments and numerical modellings so far [13-16].

The phase-space structure of the filament-arc negative ion source was investigated with 
a pepper-pot type phase space analyzer, 
and three Gaussian beam components were identified in an isolated single beamlet [17]. 
%%%%%%%%%%%%%%%%%%%%%%%%%%%%%%%%%%%%%%%%%%%%%%%%%%%%%%%%%%%   ここが大事　%%%%%%%%%%　
A backward beam calculation based on the experimentally observed phase-space structure revealed 
a non-uniform and asymmetric negative ion current profile in front of the PG aperture  [18], 
which may cause an asymmetric meniscus. 
Thus, optimization of the negative ion beam focusing is much more complicated 
than for a positive ion beam, which is basically composed of a single Gaussian beam. 

Another important finding is the oscillating behavior of the negative ion beam for the J-PARC accelerator, 
which is an RF-negative hydrogen source with an RF frequency of 2 MHz. 
The beam width oscillates at 2 MHz and a beam current oscillation at 4 MHz was also observed [19-20]. 
The oscillation at 2 MHz is understood as a direct response of the meniscus to the RF field, 
and the oscillation at 4 MHz is considered to be related to plasma production.
Because inductive electric field intensity reaches the maximum value twice in one RF cycle, 
then the density fluctuation in the ion source may have a second harmonic component. 
The RF antenna of the J-PARC ion source is located inside the vacuum vessel and in the vicinity of the plasma grid aperture, 
thus, both effects by the RF field (a direct RF field effect and a plasma production effect) were observed. 
The RF antenna configuration for ITER is different from the J-PARC source, 
and the RF antenna is located outside the vacuum vessel.
The plasma diffusion in the filter magnetic field, which has a longer time scale than RF frequency, 
may dominate the plasma dynamics near the PG, 
thus, the direct RF field is considered to be more important than the density fluctuation in the RF source for ITER.

In order to clarify the direct RF effect on beam focusing and investigate the detailed characteristics, 
beamlet dynamics were investigated  
when a controlled perturbation of the RF field was applied to the filament-arc negative ion source. 
From the viewpoint of difference from the J-PARC source, the plasma production effect was minimized in this experiment 
due to independent control between plasma production and RF perturbation.

The present study experimentally shows that the beam oscillation and 
divergence are induced by the rf electric field near the PG aperture, 
while it is demonstrated that the rf-induced divergence can be minimized 
by the proper operation of the negative ion source at the perveance 
matching condition.
The present discovery suggests that minizing the influence of RF 
electric field and the elaborate design and operation of the beam 
extraction for the perveance matching contributes to satisfying the 
requirement for ITER NBI development.
 
%%%%%%%%%%%%%%%%%%%%%%%%%%%%%%%%%%%%%%%%%%%%%%%%%%%%%%%%%
\section{Results}\label{exp}
\subsection{RF perturbation in negative ion source}
%%%%%%%%%%%%%%%%%%%%%%%%%%%%%%%%%%%%%%%%%%%%%%%%%%%%%%%%%
The experiment was carried out with a negative ion beam test stand at the National Institute for Fusion Science (NIFS-NBTS), 
and a research development negative ion source (NIFS-RNIS) was utilized, 
which is a one-third scaled negative ion source for the Large Helical Device (LHD). 
 
\begin{figure}[tb]
\includegraphics[width=0.8\textwidth]{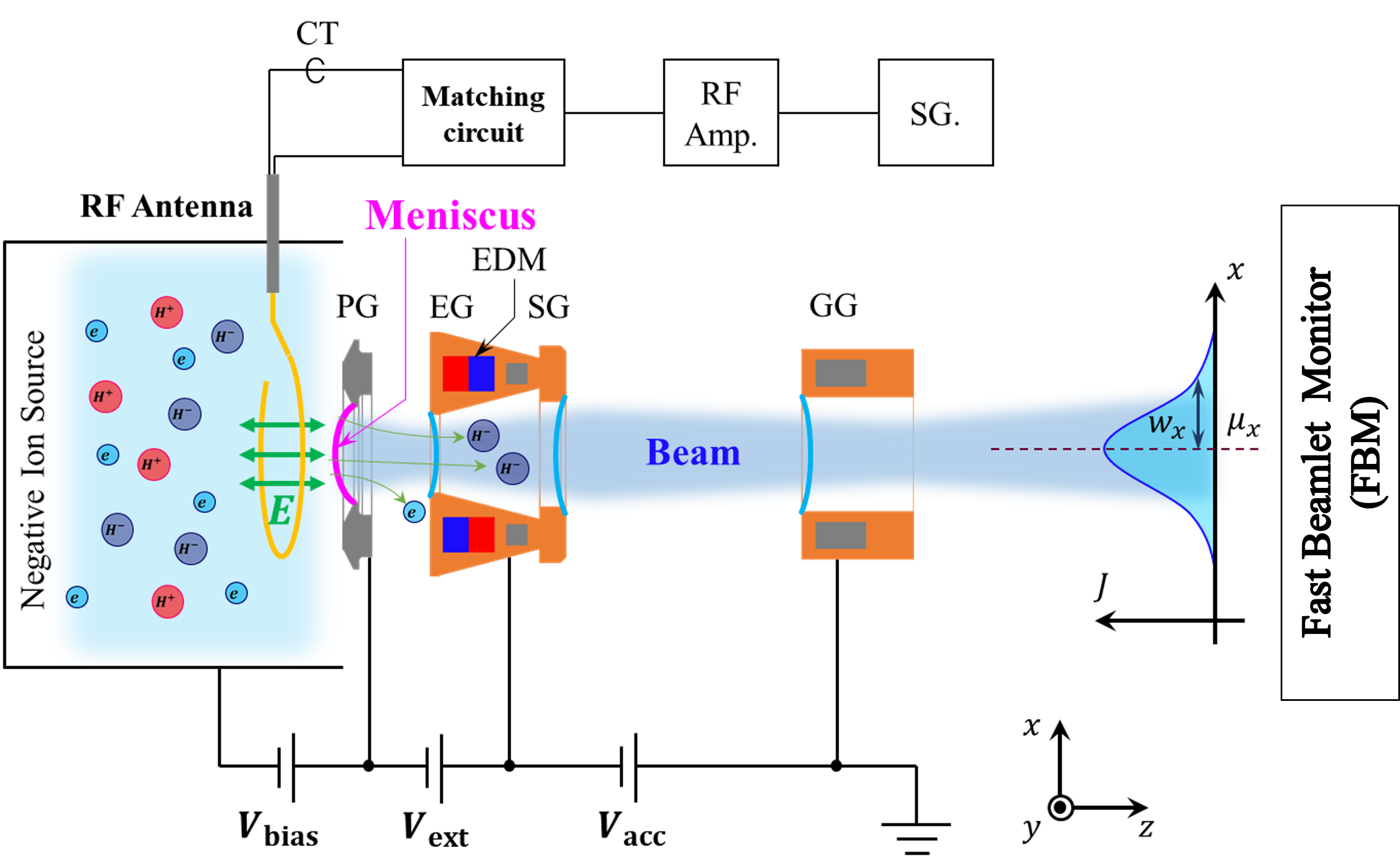}% Here is how to import EPS art
\caption{\label{fig:sche} A schematic of the experiment. }
\end{figure}

 A schematic of this experiment is shown in Fig. \ref{fig:sche}.
A Rogoski coil-type RF antenna was installed into NIFS-RNIS and the distance from the PG was around 5 mm.
The RF electric field applied by the antenna in this experiment was perpendicular to the PG, 
which is considered to be the same configuration as the RF negative ion source configuration. 
%In the negative ion source, the wavelength with a frequency of 1 MHz is much larger than the scale length of the plasma profile. 
%Therefore, the perpendicular component to the PG surface (metal surface) dominates the RF electric field.
The RF signal produced by a signal generator was amplified by a solid-state RF amplifier with a factor of $10^{6}$ (a gain of 60 dB)
and the maximum power was 1 kW, which was typically $2\%$ of the arc discharge power ($\sim 50$kW), 
resulting in negligible perturbation of the plasma density in this experiment. 
The RF power to the antenna was monitored by a current transformer (CT) with high-time resolution inserted 
between the antenna and the impedance-matching circuit. 
A PG mask was utilized to make a single isolated beamlet. 
The beamlet profile was measured with a fast beamlet monitor (FBM) with a time resolution of 25 MHz.
The FBM was installed at 0.91 m downstream from the GG. 
 
\begin{figure}[tb]
\includegraphics[width=0.99\textwidth]{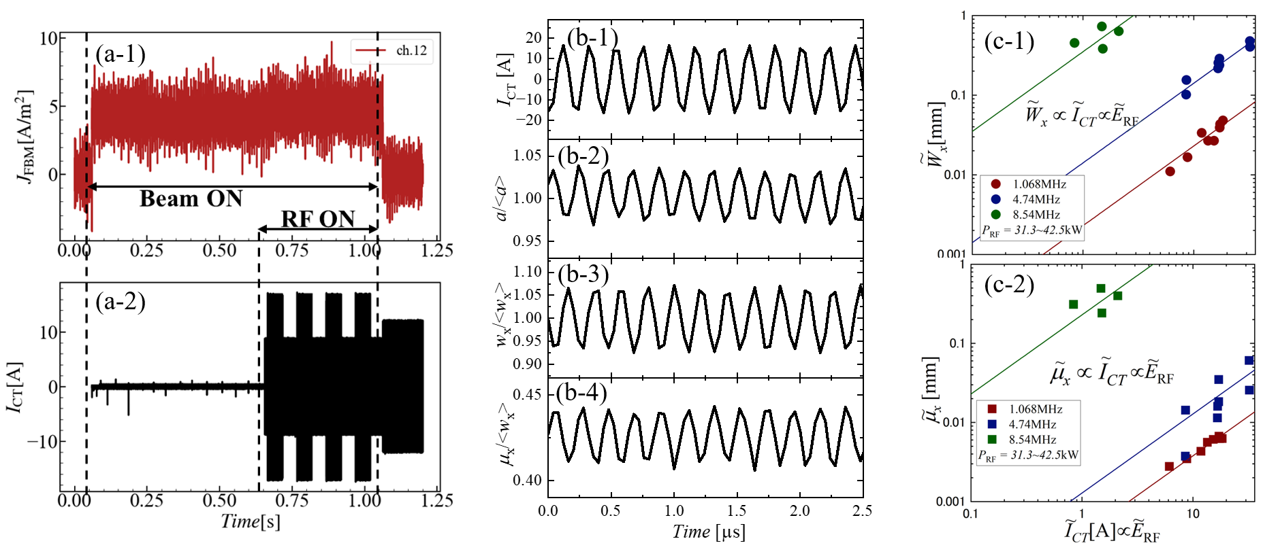}% Here is how to import EPS art
\caption{\label{fig:234} A typical waveform of (a-1) the beam current density measured 
with 12ch of FBM ($J_{\rm FBM}$) and (a-2) current of RF antenna measured with current transformer ($I_{\rm CT}$).
The power of perturbation RF is modulated.
 Time evolutions of (b-1) RF antenna current measured by the CT, 
(b-2) the peak value, (b-3) e-holding half width and (b-4) axis position of the Gaussian fitting 
to experimentally observed negative ion beam current density profile.
The bracket $< >$ means the average value in time.
RF current dependence of the oscillation amplitude of
(c-1) the beamlet width and (c-2) beamlet axisi position. 
The horizontal axis is the RF current amplitude of RF antenna ($\tilde I_{\rm CT}$) measured by the current transformer (CT), 
which could be proportional to the RF electric field ($\tilde E_{\rm RF}$).
Lines show the linear relation, that is, $\tilde{w}_{\rm x} \propto \tilde{I}_{\rm CT}$ or $\tilde{\mu}_{\rm x} \propto \tilde{I}_{\rm CT} $.
}
\end{figure}

%%%%%%%%%%%%%%%%%%%%%%%%%%%%%%%%%%%%%%%%%%%%%%%%%%%%%%%%%
%\section{Experimental Results}\label{result}
%%%%%%%%%%%%%%%%%%%%%%%%%%%%%%%%%%%%%%%%%%%%%%%%%%%%%%%%%
After the commissioning operation with caesium seeding, a negative ion beam acceleration with the perturbation RF field was carried out. 
The negative ion beam duration was 1 sec, and the perturbation RF was applied during the last one-third period of the beam.
The RF power was modulated in time to investigate the RF power dependence. 
A typical waveform of the beam and RF perturbation is shown in Fig.\ref{fig:234}-(a).

In order to investigate the beamlet behavior, a horizontal beamlet profile was analyzed with Gaussian fitting,
\begin{equation}
J(x)=a \exp \bigg[ -\frac{(x-\mu_{\rm x})^2}{w_{\rm x}^2}\bigg] +J_{\rm c},
\end{equation}
where $a$, $\mu_{\rm x}$, $w_{\rm x}$ and $J_{\rm c}$ were the peak current density, 
the position of the beamlet center, the e-holding half width of the beamlet and the offset, respectively.
The time evolution of the parameters obtained by the Gaussian fitting is shown in Fig.\ref{fig:234}-(b-1)-(b-4).
An RF with a frequency of 4.74 MHz was applied to the plasma near the PG 
and the CT signal is shown in Fig. \ref{fig:234}-(b-1). 
The oscillations with the perturbation RF frequency were observed in all three parameters, 
$a$, $w_{\rm x}$, and $\mu_{\rm x}$, which are shown in Fig.\ref{fig:234}(b-2)-(b-4).
The peak-to-peak amplitude of the beamlet width oscillation was up to $20 \%$ of the averaged beamlet width, 
which reflects a change in the beam focus.
The oscillation amplitude of the beamlet center position was up to $8 \%$ of the averaged beamlet width, 
which also directly degraded the beam focusing. 

Regarding the plasma production effect, the total current of the isolated beamlet could not be measured in this experiment 
because of the limited profile measurement in the vertical direction. 
However, the out-of-phase relation between the beamlet amplitude oscillation and the beamlet width oscillation
indicates that the total current oscillation is small. 
Another piece of evidence is that no oscillation with the second harmonic frequency was observed. 
Therefore, the plasma production effect is considered to be negligible in this experiment.

%It is also interesting that
In the case of RF positive ion sources,  these oscillations were not observed in a core region of positive ion beamlet, 
in which the perturbation RF was applied between the PG and the plasma [21]. 
This is consistent with the fact that the beam divergence angle of the RF positive ion source is almost identical 
to that of the filament-arc positive ion source.
Therefore, the beamlet oscillations observed in this experiment are considered 
to be a cause of the larger beam divergence angle of the RF negative ion source.
%This indicates that the meniscus formation mechanism for negative ion beam extraction is different from that of positive ion beam extraction.
%Therefore, 
In order to improve the beam focusing, the investigation of the dynamic response characteristics 
of the negative ion beamlet to the RF perturbation is very important.  

The RF amplitude dependence and the RF frequency dependence of the beamlet response 
were investigated and summarized in Fig.\ref{fig:234}-(c-1) and (c-2). 
Three frequencies of 1.1 MHz, 4.7 MHz and 8.5 MHz were plotted, 
and the responses of the beamlet width increased almost linearly with the antenna current.
The trend of the beamlet center position looks similar. 
The RF amplitude dependencies suggest that the meniscus responds linearly 
to the perturbation RF electric field 
because the RF electric field induced by the antenna is proportional to the RF current flowing through the antenna.
It is considered that the RF electric field was superposed to the sheath electric field, 
and the response of the meniscus to an external perturbation with a small amplitude might be linear. 
It should be emphasised that the lower frequency might give a solution to mitigate the beamlet oscillation.
In the lower frequency range from 1.1 MHz to 4.7 MHz, the oscillation amplitude of the beamlet 
seems to be proportional to the frequency (Faraday's law of induction). 
Therefore, further mitigation might be possible with lower frequency. 
In the high frequency regime from 4.7 MHz to 8.5 MHz, the frequency dependence became stronger. 
The different frequency dependence would be considered by the change of coupling between the antenna and the plasma, 
however, further investigation is necessary.

%%%%%%%%%%%%%%%%%%%%%%%%%%%%%%%%%%%%%%%%%%%%%%%%%%%%%%%%%
\subsection{perveance dependence}\label{discussion}
%%%%%%%%%%%%%%%%%%%%%%%%%%%%%%%%%%%%%%%%%%%%%%%%%%%%%%%%%

\begin{figure}[tb]
\includegraphics[width=0.99\textwidth]{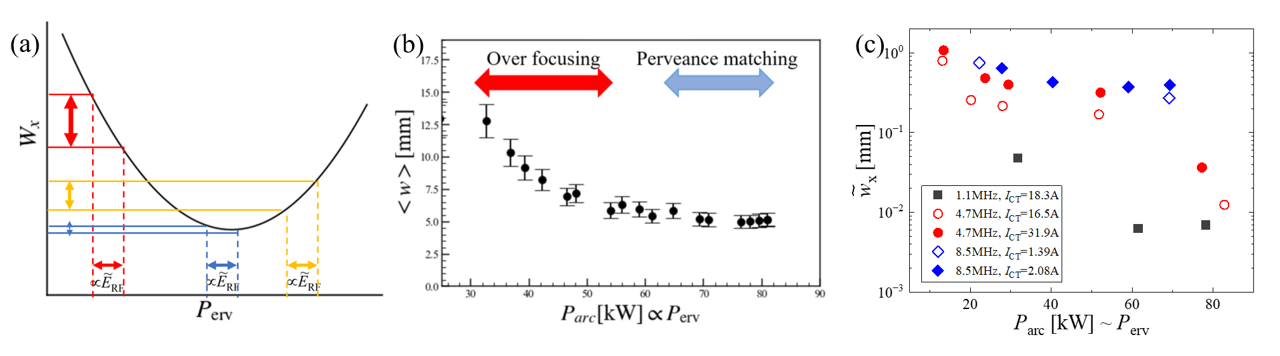}% Here is how to import EPS art
\caption{\label{fig:567} (a) A conceptual drawing of the dependence 
on the gradient of the perveance curve shown in eq. \ref{eq:model}. 
The relation between beamlet width oscillation ($\tilde W_{\rm x}$) 
and the perturbation RF electric field  ($\tilde E_{\rm RF}$)  are also shown.
(b) Arc-power dependence of the beamlet width without RF perturbation. 
The beam current is proportional to the arc power in this operational regime in the present experiment. 
Thus, it can be regarded as a perveance dependence.
(c) Arc-power dependence of the oscillation amplitude of the beamlet width.
}
\end{figure}

Here, we discuss the perveance dependence of the beamlet dynamics 
based on the experimental observations, which are shown in Fig. \ref{fig:234}-(c-1) and (c-2), 
in which $\tilde w_{\rm x} \propto \tilde E_{\rm RF}$. 
We assume that the beamlet dynamics were determined by the response of the meniscus. 
In the case of conventional perveance dependence, 
the beamlet width is determined by the balance 
between the penetration of the electric field from the beam extraction region to the plasma 
and the Debye shielding effect of the ion source plasma. 
When an external perturbation is applied and the meniscus responds linearly, 
the change in beamlet width should depend on the gradient of the perveance curve,
\begin{equation}\label{eq:model}
\tilde w_{\rm x} \propto \tilde E_{\rm RF} \bigg( \frac{\partial <w_{\rm x}>}{\partial P_{\rm erv}} \bigg),
\end{equation} 
where $\tilde E_{\rm RF}$ and $P_{\rm erv}$ are the perturbation RF electric field amplitude and the perveance 
($P_{\rm erv}=I_{\rm beam}/V_{\rm extract}^{1.5}$, 
where $I_{\rm beam}$ is the beam current and $V_{\rm extract}$ is the extraction voltage), respectively.
The idea of eq. \ref{eq:model} is sketched in Fig. \ref{fig:567}-(a). 
It should be noted that the $\tilde E_{\rm RF}$ corresponds to the oscillation of the meniscus 
and is not $\tilde P_{\rm erv}= {\tilde I_{\rm beam}}/{V_{\rm extract}^{1.5}}$.

\begin{figure}[tb]
\includegraphics[width=0.45\textwidth]{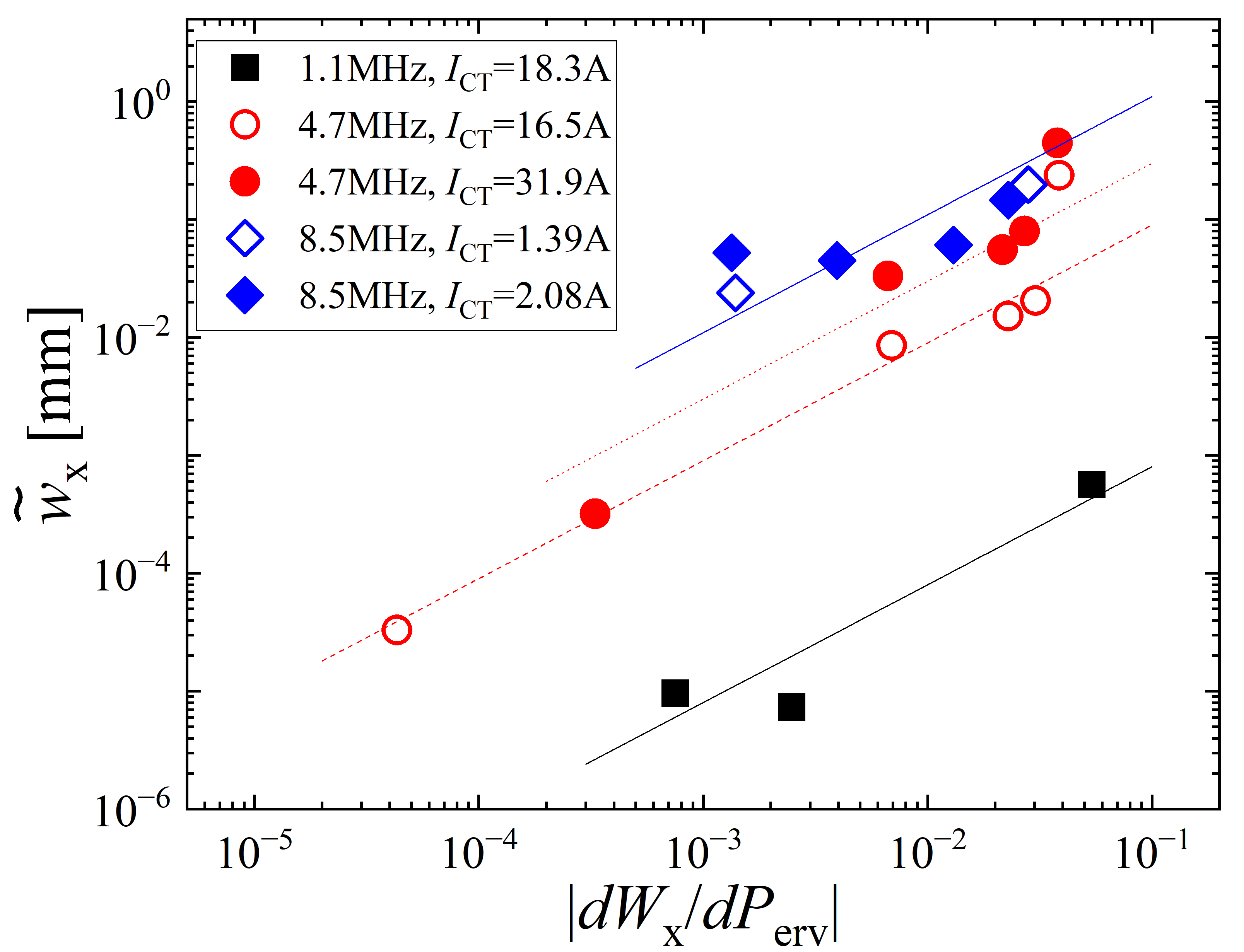}
\caption{\label{fig:gradient} Dependence of the oscillation amplitude of the beamlet width 
on the gradient of the perveance curve. }
\end{figure}

Figure \ref{fig:567} (b) shows the perveance dependence of the beamlet width without the RF perturbation. 
One can see the perveance matching and over-focus regions due to the weak Debye shielding effect. 
A similar dependence can be seen in the oscillation amplitude of the beamlet width 
when RF perturbation is applied, which is shown in Fig. \ref{fig:567} -(c). 
In order to investigate the validity of the hypothesis shown in eq. \ref{eq:model}, 
the amplitude of the beamlet width oscillation is compared with the gradient of the perveance curve 
${\partial <w_{\rm x}>}/{\partial P_{\rm erv}}$, which is shown in Fig. \ref{fig:gradient}.
One can see the almost linear relation between the oscillation amplitude of the beamlet width 
and the gradient of the perveance curve when the RF power is fixed. 
This indicates that the oscillation of the beamlet width can be understood 
by the direct response of the sheath boundary to the perturbation RF electric field, 
which is modelled by eq. \ref{eq:model}. 
The RF frequency dependence is also clearly seen, 
and the lower RF frequency provides the smaller amplitude of the beamlet width oscillation.
From the viewpoint of the beam focusing, 
it should be noted that the oscillation of the beamlet width can be minimized 
at the perveance matched condition, where the gradient of the perveance curve is zero 
(${\partial <w_{\rm x}>}/{\partial P_{\rm erv}} \sim 0$).

%%%%%%%%%%%%%%%%%%%%%%%%%%%%%%%%%%%%%%%%%%%%%%%%%%%%%%%%%
%\section{Concluding remarks}\label{conclusion}
%%%%%%%%%%%%%%%%%%%%%%%%%%%%%%%%%%%%%%%%%%%%%%%%%%%%%%%%%

In this study, the direct effect of the RF field on beamlet focusing has been experimentally investigated. 
Oscillations of the beamlet width 
and of the beamlet center position were observed 
when the perturbation RF was applied to the plasma near the PG aperture. 
In comparing the RF positive ion sources, the larger beam divergence in RF negative ion sources is 
attributable to the oscillation of the meniscus and resulting in the oscillations of the beamlet width and position.
The physics mechanism of the different responses of the meniscus is still an open issue.

The perveance dependence of the oscillation of the beamlet width was demonstrated 
and it was found that the beamlet width oscillation could be reduced 
at low RF frequency and at the perveance matching condition.
It should be noted that the importance of the perveance matching is more significant in the RF negative ion source 
than in the filament-arc negative ion source.

The oscillation of the beamlet center is considered to be caused 
by the oscillation of the asymmetric meniscus.
Although the PG aperture geometry is circularly symmetric, 
the magnetic field for electron deflection may break 
the symmetry of the particle dynamics in front of the PG aperture [18, 22-24], 
where the ion Larmor radius is in the same order as the aperture diameter. 
In order to mitigate the oscillation of the beamlet center, 
some modifications of the PG aperture geometry to cancel the asymmetry of the meniscus seem to make sense. 

Quantitative evaluation of the beamlet response to the RF field 
in an ITER-relevant configuration of the RF negative ion source 
and a dedicated design study of the PG aperture configuration is very important research 
to improve the beam focusing of RF negative ion sources 
and expand the application of the negative ion beams.

%%%%%%%%%%%%%%%%%%%%%%%%%%%%%%%%%%%%%%%%%%%%%%%%%%%%%%%%%
\section{Methods}
\subsection{NIFS-NBTS}\label{nbts}
%%%%%%%%%%%%%%%%%%%%%%%%%%%%%%%%%%%%%%%%%%%%%%%%%%%%%%%%%
This study was carried out using the negative ion beam test stand at the National Institute for Fusion Science (NIFS-NBTS), 
in which the full-size negative ion source for the LHD plasma experiment can be operated with full specification of the negative ion source.

%%%%%%%%%%%%%%%%%%%%%%%%%%%%%%%%%%%%%%%%%%%%%%%%%%%%%%%%%
\subsection{NIFS-RNIS}\label{rnis}
\begin{figure}[tb]
\includegraphics[width=0.99\textwidth]{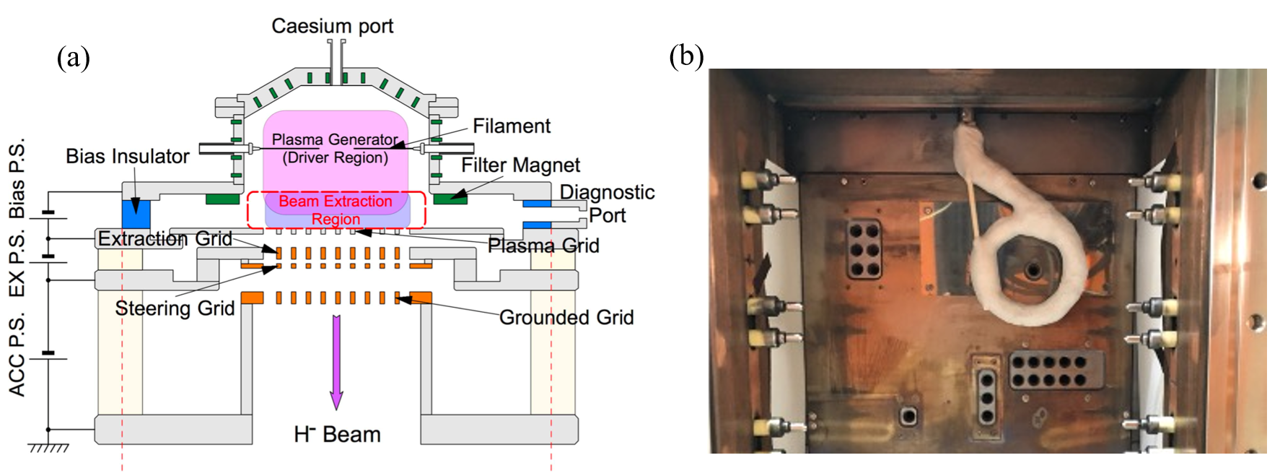}
\caption{\label{fig:rnis} (a) Drawing of the horizontal cross-section of RNIS [25]. (b) Photograph of the inside of the RNIS with installed RF antenna and PG mask. The beamlet produced through the isolated PG aperture at the center of the RF antenna was measured with FBM in this experiment. }
\end{figure}

%%%%%%%%%%%%%%%%%%%%%%%%%%%%%%%%%%%%%%%%%%%%%%%%%%%%%%%%%
The NIFS-RNIS is a research and development of negative ion sources (see Fig. \ref{fig:rnis}-(a)), which is operated at NIFS-NBTS. 
The size of NIFS-RNIS used in this study was 1/3 of the negative ion source for the LHD plasma experiment.
The inner volume of arc chamber is $700 \times 350 \times 220$ mm. 
The beam accelerator consists of two segments of a four-grid system. 
The beam extraction area of each segment is approximately $250 \times 250$ mm. 
The accelerator configuration was almost identical to that of the negative ion source of the LHD, 
a circular multi-aperture for the plasma grid (PG), an extraction grid (EG) with an electron deflection magnet, a steering grid (SG) 
and a multi-slot type grounded grid (GG). 
Negative ions were produced mainly by the surface process on a caesium-seeded PG surface.
In order to monitor a single isolated beamlet, the PG make plate was mounted onto the PG. 

%%%%%%%%%%%%%%%%%%%%%%%%%%%%%%%%%%%%%%%%%%%%%%%%%%%%%%%%%
\subsection{RF system}\label{rfsystem}
%%%%%%%%%%%%%%%%%%%%%%%%%%%%%%%%%%%%%%%%%%%%%%%%%%%%%%%%%
The RF perturbation field was applied with an RF antenna installed in NIFS-RNIS (see Fig. \ref{fig:rnis}-(b)). 
The distance from the PG was around 5 mm. 
The direction of the RF electric field is perpendicular to the PG. 
RF signal with the power level of mW was produced by a signal generator. 
The RF signal was amplified with a factor of $10^6$ (a gain of $60$ dB) using an RF power amplifier (R$\&$K, CA009251-5959R). 
The amplifier's maximum power is $1$ kW, and its frequency range is from 9 kHz to 250 MHz. 
In this study, the maximum RF power of 1 kW and the frequency range of from 1.1 MHz to 8.5 MHz were utilized.
A $\pi$-type RF matching circuit was developed and used for impedance matching of RF power to the RF antenna.
The RF power to the RF antenna was monitored by a current transformer (CT) with high-time resolution, 
which was inserted between the antenna and the impedance-matching circuit.
The RF electric field generated by the RF antenna was measured before installation into the NIFS-RNIS. 
The maximum RF electric field is roughly 3 kV/m, with 1 kW of RF power to the antenna and 1MHz of frequency.

%%%%%%%%%%%%%%%%%%%%%%%%%%%%%%%%%%%%%%%%%%%%%%%%%%%%%%%%%
\subsection{Fast Beamlet Monitor}\label{fbm}
%%%%%%%%%%%%%%%%%%%%%%%%%%%%%%%%%%%%%%%%%%%%%%%%%%%%%%%%%
The beamlet profile was measured with a fast beamlet monitor (FBM) installed at  $0.91$ m downstream from the GG.
The FBM consists of $8 \times 4$ ch array of current monitors in horizontal and vertical directions, respectively. 
The bias voltage of $+ 70$ V  was applied to the collector grid for suppression of the secondary electron current so the absolute values of the negative ion beam current density could be measured.
The time resolution of the FBM diagnostic is 25 MHz; therefore, the beamlet's responses to the RF perturbation can be monitored.

\section{Data Availability}

The data that support the findings of this study are available from the corresponding author upon request.


\begin{enumerate}
%\begin{thebibliography}{99}
    %\setlength{\itemindent}{-\leftmargin}
    \renewcommand{\labelenumi}{[\theenumi]}
    
    \item R.S. Hemsworth, D. Boilson, et al., New J. Phys., 19, 025005 (2017).
    \item U. Fantz, S. Briefi, A. Heiler, C. Wimmer and D. Wunderlich, Frontier in Physics, 9, 709651 (2021). 
    \item M. Barbisan, B. Zaniol, et al., Plasma Phys. Control., Fusion, 63, 125009 (2021). 
    \item A. Hurbatt, F. Bonomo, et al., AIP Advances, 11, 025330 (2021).
    \item L. Hu et al., Fusion Eng. Des., 54, 321 (2001).
    \item K. Tsumori, et al., Rev. Sci. Instrum., 79, 02C107 (2008).
    \item K. Tsumori, et al., Rev. Sci. Instrum., 81, 02B117 (2010).
    \item P. Veltri, et al., Rev. Sci. Instrum., 87, 02B908 (2016).
    \item N.D. Harder, et al., Proceedings of the 29th IAEA-FEC, IAEA-CN-316-1882, 16-21 October 2023, London, United Kingdom.
    \item C. Wimmer, et al., The 20th International Conference on Ion Sources, 17-22 September 2023.
    \item U. Fantz, et al.,  Proceedings of the 29th IAEA-FEC, IAEA-CN-316-1794, 16-21 October 2023, London, United Kingdom.
    \item P. Veltri, N. Den Harder, et al., 8th International Symposium on Negative Ions, Beams and Sources, Padova, Italy, Oct. 2-7, 2022.
    \item S. Mochalskyy, U. Fantz, et al., Nucl. Fusion, 56, 106025 (2016).
    \item T. Kalvas, O. Tarvanien, et al., Rev. Sci. Instrum., 81, 02B703 (2010).
    \item A. Ueno et al., Rev. Sci. Instrum., 91, 033312 (2020).
    \item M. Ichikawa, et al., AIP Conf. Proc.1869, 030024 (2017).
    \item Y. Haba, K. Nagaoka, et al., New J. Phys. 22, 023017 (2020).
    \item M. Kisaki, K. Nagaoka, et al., Nucl. Fusion, 62, 106031 (2022).
    \item T. Shibata, et al., AIP Conf. Proc. 2373, 050002 (2021).
    \item M. Wada et al., 8th International Symposium on Negative Ions, Beams and Sources, Padova, Italy, Oct. 2-7, 2022.
    \item K. Takahashi, et al., New J. Phys., 21, 093043 (2019).
    \item R. Gutser, D. Wunderlich U. Fantz and NNBI-Team, Plasma Phys. Control. Fusion, 51, 045005 (2009). 
    \item G. Fubiani, L. Garrigues and J.P. Boeuf, Phys. Plasmas, 25, 023510 (2018). 
    \item K. Miyamoto, K. Nagaoka, A. Hatayama, et al., J. Phys. Conf. Series, 2244, 012040 (2022).
    \item K. Tsumori and M. Wada, New J. Phys., 19, 045002 (2017).

\end{enumerate}
%\end{thebibliography}

\section{Acknowledgements}
\begin{acknowledgments}
One of the authors (K.N.) would like to thank Dr. W. Kraus (Max Planck Institute for Plasma Physics), 
Dr. T. Sasaki (Nagaoka University of Technology), 
Dr. K. Shinto, Dr. T. Shibata (KEK) and Dr. M. Wada (Doshisya University) 
for the fruitful discussions and engineering staff of NBI group at NIFS 
for their support of the experimental setup and the beam operations. 
This research was partially supported by NIFS(NIFS23KIIR023) and JSPS KAKENHI (17H03002, 18KK0080). 
\end{acknowledgments}

\section{Author contributions}
K. N. proposed this experimental study with discussions with U. F., K. T. and M. O.. 
K. N. also conducted the experiments and implemented the FBM. 
H. N. and R. N. designed and constructed the RF antenna and impedance matching circuit. 
T. H. analyzed the beamlet oscillation based on the FBM data. 
K. T. and K. I. conducted the negative ion beam operation with good beam focusing conditions. 
M. O. discussed the meniscus dynamics responding to applied RF perturbation based on the FBM data.
M .K. analyzed the perveance dependence of the beamlet response. 
K. M. discussed the perveance dependence modelling for the beamlet response. 
K. T. designed the total RF system of the experimental study.
U. F. discussed the frequency dependence of the beamlet response. 
All authors participated in the experimental planning and discussions of the data analyses and contributed to improving the manuscript.

\section{Competing interests}
The authors declare no competing interest.

%\section{Additional information}

\end{document}